\documentclass{article}

\usepackage{PRIMEarxiv}

\usepackage[utf8]{inputenc} 
\usepackage[T1]{fontenc}    
\usepackage{hyperref}       
\usepackage{url}            
\usepackage{booktabs}       
\usepackage{amsfonts}       
\usepackage{nicefrac}       
\usepackage{microtype}      
\usepackage{lipsum}
\usepackage{fancyhdr}       
\usepackage{graphicx}       
\graphicspath{{media/}}     
\usepackage{amsmath}
\usepackage{epstopdf}

\title{Reinforcement Learning Based Prediction of PID Controller Gains for Quadrotor UAVs}

\author{Serhat Sönmez$^{*\dagger}$, Luca Montecchio$^*$, Simone Martini$^*$, Matthew J. Rutherford, \\ \textbf{Alessandro Rizzo, Margareta Stefanovic, and Kimon P. Valavanis} \\ \\
* Co-First Authors, Equal Contribution\\
$\dagger$ Corresponding Author \\
{\tt\small serhat.sonmez@du.edu},
{\tt\small luca.montecchio@polito.it},
{\tt\small simone.martini@du.edu},\\
{\tt\small matt.rutherford@du.edu},
{\tt\small alessandro.rizzo@polito.it},\\
{\tt\small margareta.stefanovic@du.edu},
{\tt\small kimon.valavanis@du.edu}}

\begin{document}
\maketitle

\begin{abstract}
A reinforcement learning (RL) based methodology is proposed and implemented for online fine-tuning of PID controller gains, thus, improving quadrotor effective and accurate trajectory tracking. The RL agent is first trained offline on a quadrotor PID attitude controller and then validated through simulations and experimental flights. RL exploits a Deep Deterministic Policy Gradient (DDPG) algorithm, which is an off-policy actor-critic method. Training and simulation studies are performed using Matlab/Simulink and the UAV Toolbox Support Package for PX4 Autopilots. Performance evaluation and comparison studies are performed between the hand-tuned and RL-based tuned approaches. The results show that the controller parameters based on RL are adjusted during flights, achieving the smallest attitude errors, thus significantly improving attitude tracking performance compared to the hand-tuned approach.
\end{abstract}


\section{Introduction}
Unmanned aerial vehicles (UAVs) have experienced tremendous growth over the past two decades, and they have been utilized in diverse civilian and public domain applications like power line inspection \cite{martinez2014towards}, monitoring mining areas \cite{ren2019review}, wildlife conservation and monitoring \cite{olivares2015towards}, border protection \cite{bassoli2019virtualized}, infrastructure and building inspection \cite{carrio2016ubristes}, and precision agriculture \cite{li2016real}, among others. Multirotor UAVs, particularly quadrotors, have become the most widely used platforms due to their vertical take-off and landing (VTOL) capabilities, efficient hovering, and overall flight effectiveness. 

 Although several conventional control techniques have been developed and tested effectively (via simulations and in real time) for quadrotor navigation and control, recently, learning-based algorithms and techniques have gained significant momentum because they improve quadrotor modeling and subsequently navigation and control. The learning-based methodology offers alternatives to parameter tuning and estimation, learning, and understanding of the environment. Representative published surveys on developing and adopting machine learning (ML), deep learning (DL), or reinforcement learning (RL) algorithms for UAV modeling and control include \cite{carrio2017review},  \cite{polydoros2017survey}, \cite{choi2019unmanned}, \cite{azar2021drone}, \cite{brunke2022safe}, while the recently completed survey in \cite{sönmez2024survey} focuses on multirotor navigation and control based on online learning.

This paper focuses on the applicability of the RL-based PID controller tuning method, initially derived in \cite{sonmez2024reinforcement}, on a real quadrotor that is used as the test platform. However, the X configuration is followed instead of the conventional + one, and a circular trajectory is used for tracking instead of a helix trajectory (that was previously tested). Therefore, the agent is retrained for the new quadrotor configuration and task execution. The followed approach is also suitable to test PD controllers, and it is also applicable when using other control strategies.

This study makes four primary contributions. First, a novel application of the DDPG-based reinforcement learning algorithm is introduced for fine-tuning the inner-loop gains of PD controllers, enhancing adaptability and performance in UAV trajectory tracking. The proposed approach eliminates the need for a full system model, including drag effects, gyroscopic effects, or external/environmental disturbances such as wind and sensor noise, during the training phase, thereby simplifying implementation. Second, the methodology is validated through a comprehensive evaluation pipeline comprising simulation, Hardware-In-The-Loop (HIL) testing, and real-world outdoor experiments. During real-world flights, the controller gains are fine-tuned dynamically to account for external and environmental disturbances. Third, critical challenges such as GPS inaccuracies and the omission of certain physical effects during training are addressed, providing insights for future research and practical UAV applications. Finally, the multi-stage validation demonstrates the robustness and scalability of the reinforcement learning-enhanced controller across various conditions, including environmental disturbances and complex trajectories. These contributions advance the understanding and implementation of reinforcement learning-based control strategies for UAVs, bridging the gap between simulation and real-world operations.

The rest of the paper is organized as follows: Section~\ref{sec2} provides notation and background information related to the quadrotor mathematical model, the implemented PD controller with feedback linearization, and RL.  Section~\ref{sec3} introduces the training phase in MATLAB/Simulink environment. Section~\ref{sec4} dives into the RL agent environment setup and its deployment to real hardware. The training phase, numerical simulations, and experimental results are provided in Section~\ref{sec5}. Finally, Section~\ref{sec6} concludes the paper.  

\section{Notation \& Background Information}\label{sec2}
\subsection{Quadrotor Mathematical Model}
Recall that given two vectors $a = \left[a_1, a_2, a_3 \right]^\top$ and $b = \left[b_1, b_2, b_3 \right]^\top$, the matrix $S(a)$ denotes the skew-symmetric matrix
\begin{equation}
S(a) =
\begin{bmatrix}
0 & -a_3 & a_2 \\
a_3 & 0 & -a_1 \\
-a_2 & a_1 & 0
\end{bmatrix}
\end{equation}
for which the following relation holds $S(a)b = a \times b$.

Next, the quadrotor mathematical model is derived considering the ``$\times$" configuration, as opposed to the ``$+$" configuration adopted in \cite{sonmez2024reinforcement}. 

The quadrotor Newton-Euler (N-E) equations of motions are given as follows:
\begin{subequations}
\begin{equation}
\ddot p = -g{\hat z} + \frac{1}{m}T{{\mathop{\rm R{\hat z}}\nolimits}} \label{FullNEMotionEquation2}
\end{equation}
\begin{equation}
\dot R = RS({\omega ^B}) \label{FullNEMotionEquation3}
\end{equation}
\begin{equation}
{I_f}{{\dot \omega }^B} =  - {\omega ^B} \times ({I_f}{\omega ^B}) + M_B \label{FullNEMotionEquation4}
\end{equation}
\end{subequations}
Let $p = (x, y, z)^T$ express the position of the body-fixed frame $B$ with respect to the inertial frame $E$. Gravitational acceleration and mass are defined by $g$ and $m$, respectively. The unit vector along the $z$-axis is represented by $\hat{z} = (0, 0, 1)^T$.
The rotation matrix $R \in SO(3)$ maps vectors from the body-fixed frame to the inertial reference frame and it is parameterized using Euler angles $\eta = \left[ \varphi,\theta,\psi \right]^\top$ as in \cite{martini2024correction}
\begin{equation}
R =
\begin{bmatrix}
c_\theta c_\psi & s_\phi s_\theta c_\psi - c_\phi s_\psi & c_\phi s_\theta c_\psi + s_\phi s_\psi \\
c_\theta s_\psi & s_\phi s_\theta s_\psi + c_\phi c_\psi & c_\phi s_\theta s_\psi - s_\phi c_\psi \\
-s_\theta & s_\phi c_\theta & c_\phi c_\theta
\end{bmatrix}\label{eq:rotation_matrix}
\end{equation}
where $c_\alpha = \cos \alpha$, $s_\alpha = \sin \alpha$, $\alpha \in \{\phi, \theta, \psi\}$. $\omega^B = \left[ p, q, r \right]^\top$ represents angular velocities in the body-fixed frame which relationship to the Euler rates is expressed by 
\begin{equation}
 \begin{bmatrix}
\dot{\varphi} \\
\dot{\theta} \\
\dot{\psi} \\
\end{bmatrix}
=
\underbrace{\begin{bmatrix}
1 & \frac{s(\varphi)s(\theta)}{c(\theta)} & \frac{c(\varphi)s(\theta)}{c(\theta)} \\
0 & c(\varphi) & -s(\varphi) \\
0 & \frac{s(\varphi)}{c(\theta)} & \frac{c(\varphi)}{c(\theta)} 
\end{bmatrix}}_{W^{-1}}
\begin{bmatrix}
p \\
q \\
r \\
\end{bmatrix}
\end{equation}
where $W$ is defined as in \cite{martini2024correction}. $I$ is the symmetric and positive definite inertia matrix which is computed with respect to the airframe's center of mass and expressed in the body-fixed frame. Finally, $M_B = [M_p, M_q, M_r]^T$ is the external torque expressed in the body-fixed frame. 

Following \cite{martini2024design}, the quadrotor's model forcing terms are the results of a linear combination of each propeller thrust $T_i$
\begin{subequations}
\begin{align}
    T &= ({T_1} + {T_2} + {T_3} + {T_4})\label{eq:T}\\
    M_p &= \frac{{\sqrt 2 }}{2}l({T_2} + {T_3} - {T_1} - {T_4})\label{eq:tau_p}\\
    M_q &= \frac{{\sqrt 2 }}{2}l({T_1} + {T_3} - {T_2} - {T_4})\label{eq:tau_q}\\
    M_r &= \frac{{{c_D}}}{{{c_L}}}({T_1} + {T_2} - {T_3} - {T_4})\label{eq:tau_r}
\end{align}
\end{subequations}
where $l$, $c_D$ and $c_L$ denote the quadrotor arm length, the linear friction coefficient and angular friction coefficient, respectively.
The thrust $T_i$ is obtained through the spinning of the $i$-th propeller and controlled by a PWM signal to be sent to the electronic speed controller (ESC) following the control interface presented in \cite{martini2024design}. To this end, it is desirable to express the forcing signals in terms of adimensional virtual control inputs $\tau_T, \tau_R, \tau_P$, and $\tau_Y$ which are dependent on the maximum input thrust $T_{max}$ according to the following relations
\begin{subequations}
\begin{align}
    T &= 4T_{max}(\tau_T -1) \label{eq:T_virtual}\\
    M_p \nicefrac{2}{l\sqrt 2} &= -4T_{max} \tau_R \label{eq:tauR_virtual}\\
    M_q \nicefrac{2}{l\sqrt 2} &= -4T_{max} \tau_P \label{eq:tauP_virtual}\\
    M_r \nicefrac{c_{L}}{c_{D}} &= 4T_{max} \tau_Y \label{eq:tauY_virtual}
\end{align}
\end{subequations}

\subsection{Quadrotor Position and Attitude Controller}

Quadrotors are treated as underactuated nonlinear systems since position and attitude control (six generalized coordinates) is achieved through the four propellers' input thrust. A hierarchical control architecture has been followed for trajectory tracking tasks \cite{l2018introduction} that is composed of an outer position control loop and an inner attitude control loop, respectively.
\subsubsection{Outer-Loop Control}\label{sec:outer_loop}
The Outer-Loop control law is formulated independently for the $z$ axis and $x$, $y$ axis respectively. The control law is achieved by inverting the position dynamics of \eqref{FullNEMotionEquation2} and has been slightly modified with respect to the one proposed in \cite{martini2024design,Tesi_Davide} since the drag effects have not been considered. 
To this end, the altitude control law is designed as
\begin{equation}
    \tau_{T}=1-\frac{m}{4T_{m a x}(\cos\varphi)(\cos\theta)}(-g+v_{z})\label{eq:tau_T_virtual}
\end{equation}
with $v_{z}$ being the altitude virtual control law to be designed.
Additionally, the desired Euler angles to achieve $x, y$-axis position control are computed as
\begin{equation}
    \left[ \begin{array}{l}
\varphi_r \\
\theta_r 
\end{array} \right] = {{F_B^*}^{ - 1}}{V_{xy}}
\end{equation}
with
\begin{equation}
F_B^* =  - \frac{{4{T_{\max }}({\tau _T} - 1)}}{m}\left[ {\begin{array}{*{20}{c}}
{\sin \psi }&{\cos \psi }\\
{ - \cos \psi }&{\sin \psi }
\end{array}} \right]
\end{equation}
and $V_{xy} = \left[v_x,v_y \right]^{\top}$ being the $x,y$-position virtual control law to be designed.

The Outer-Loop outputs are $\tau_T$, $\varphi_r$, $\theta_r$. The $\tau_T$ goes to the PWM conversion block while $\varphi_r$ and $\theta_r$, along with the desired yaw angle $\psi_r$, will be used as reference for the Inner-Loop.

The proposed control laws are formulated as follow:
\begin{subequations}
\begin{align}
    e_{\dot x}&=k_{P1,x}(x_r-x)- \dot x\\
    e_{\dot y}&=k_{P1,y}(y_r-y)- \dot y\\
    e_{\dot z}&=k_{P1,z}(z_r-z)- \dot z\\
    v_{x}&=k_{P2,x} e_{\dot x} + k_{D,x}\dot e_{\dot x}\\
    v_{y}&=k_{P2, y} e_{\dot y} + k_{D,y}\dot e_{\dot y}\\
    v_{z}&=k_{P2,z} e_{\dot z}
\end{align}
\end{subequations}
Given the symmetry of the quadrotor system, the controller gains related to the $x$ and $y$ positions are set such that  $k_{P1,x}=k_{P1,y}=k_{P1,xy}$, $k_{P2,x}=k_{P2,y}=k_{P2,xy}$, $k_{D,x}=k_{D,y}=k_{D,xy}$.
\subsubsection{Inner-Loop}\label{sec:inner_loop}
The Inner-Loop control law is designed by enhancing with feedback linearization the linear attitude controller presented in the Mathworks documentation regarding the UAV Toolbox Support Package for PX4 Autopilots \cite{HIL_Mathworks}.
The feedback linearization strategy is used to exactly linearize the quadrotor nonlinear dynamics and therefore guarantee stability away from the equilibrium points. The following implementation exploit the Euler-Lagrange modeling formulation which, in its correct form, can be used interchangeably to the N-E mathematical model as shown in \cite{martini2024correction}
\begin{equation}
\label{eq:feed_law}
    M' = W^{-T} [B v + C \dot{\eta}]
\end{equation}
where $B$ and $C$ are two $3\times3$ matrices defined as in \cite{martini2022euler} and $v= \left[v_{\varphi},v_{\theta},v_{\psi} \right]^{\top}$ is a virtual control signal resulting from the inner-loop PD controller presented below.
Compared to the one presented in \cite{HIL_Mathworks}, the inner-loop PD controller is designed using Euler rates instead of angular velocities and it's formulated as follows:

\begin{subequations}
\begin{align}
    e_{\dot \varphi}&=k_{P1,\varphi}(\varphi_r-\varphi)- \dot \varphi\label{eq:law_phi}\\
    e_{\dot \theta}&=k_{P1,\theta}(\theta_r-\theta)- \dot \theta\\
    e_{\dot \psi}&=k_{P1,\psi}(\psi_r-\psi)- \dot \psi\\
    v_{\varphi}&=k_{P2,\varphi} e_{\dot \varphi} + k_{D,\varphi}\dot e_{\dot \varphi}\\
    v_{\theta}&=k_{P2, \theta} e_{\dot \theta} + k_{D,\theta}\dot e_{\dot \theta}\\
    v_{\psi}&=k_{P2,\psi} e_{\dot \psi} \label{eq:law_phi2}
\end{align}
\end{subequations}
where $k_{P1,\varphi}$, $k_{P1,\theta}$, $k_{P1,\psi}$, $k_{P2,\varphi}$, $k_{P2,\theta}$, $k_{P2,\psi}$, $k_{D,\varphi}$, and $k_{D,\theta}$ are controllers gains to be tuned. Given the symmetry of the quadrotor system, the controller gains related to the roll and pitch angles are set such that  $k_{P1,\varphi}=k_{P1,\theta}=k_{P1,\varphi \theta}$, $k_{P2,\varphi}=k_{P2,\theta}=k_{P2,\varphi \theta}$, 
$k_{D,\varphi}=k_{D,\theta}=k_{D,\varphi \theta}$.


\subsection{Reinforcement Learning} 
\label{subsec: reinforcement learning}

\begin{figure}[b]
\centerline{\includegraphics[width=0.5\columnwidth]{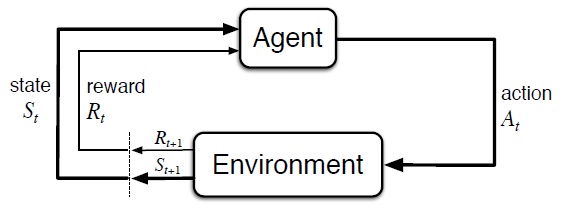}}
\caption{Interaction scheme between the agent and environment \cite{sutton2018reinforcement}.}
\label{RL_interaction}
\end{figure}

RL centers on training an agent that decides about taking actions by maximizing a long-term benefit through trial and error. RL is generally described by a Markov Decision Process (MDP). The agent-environment interaction in an MDP is illustrated in Fig.~\ref{RL_interaction}. Agent, environment, and action represent the controller, controlled system, and control signal in engineering terms, respectively \cite{sutton2018reinforcement}.

A DDPG algorithm is adopted and it is designed to handle high-dimensional action spaces. It is an off-policy actor-critic algorithm that works based on the expected gradient of the action-value function that is given by  
\begin{equation} \label{eq: action-state function}
    q_\pi(s,a) = \mathbb{E}_\pi \left[\sum_{k=0}^{\infty}\gamma^{k}R_{t+k+1} | S_{t} = s, A_{t} = a \right]
\end{equation}
where $q_\pi(s,a)$ denotes the action-value function for policy $\pi$ at state \textit{s} and action \textit{a}. $\mathbb{E}_{\pi}[\cdot]$ represents the expected value under policy $\pi$. $\sum_{k=0}^{\infty}\gamma^{k}R_{t+k+1}$ is the sum of discounted future rewards starting from time \textit{t} in state \textit{s} and represents the expected discounted return, and $\gamma$ is the discount rate, $0 \leq \gamma \leq 1$ \cite{sönmez2024survey}\cite{sutton2018reinforcement}\cite{bilgin2020mastering}.

The DDPG algorithm finds a deterministic target policy using an exploratory behavior policy. Thus, it outputs a specific action rather than a probability distribution over actions. 

The actor-critic approach includes both a value function-based and a policy search-based method. While the actor refers to the policy search-based method and chooses actions in the environment, the critic refers to the value function-based method and evaluates the actor using the value function.

Two different state spaces $S$, action space $A$, and reward functions $r$, are used because of controller parameter tuning and estimation. The dimensions of the state space change, while the dimensions of the action space remain the same. This is further explained in the next section along with the reward function differences. Moreover, the next section details the proposed method to tune and estimate controller parameters.

\section{Proposed Training Phase Methodology} \label{sec3}

It is essential to properly adjust controller parameters as this directly impacts performance. Given a PD controller, an excessively high proportional gain may lead to overshooting and a very high derivative gain may lead to instabilities. Thus, achieving an 'optimal balance' among the three gains is crucial to ensure a PD controller's accurate and stable response (when applied to different dynamic systems). 

The MATLAB/Simulink environment has been used for the presented simulation studies. While Simulink offers a good tool for PD controller tuning, challenges arise due to difficulties in linearizing the plant (using Simulink). Consequently, manual tuning of PD parameters is performed.

\begin{figure*}[h!]
\centerline{\includegraphics[width=0.9\columnwidth]{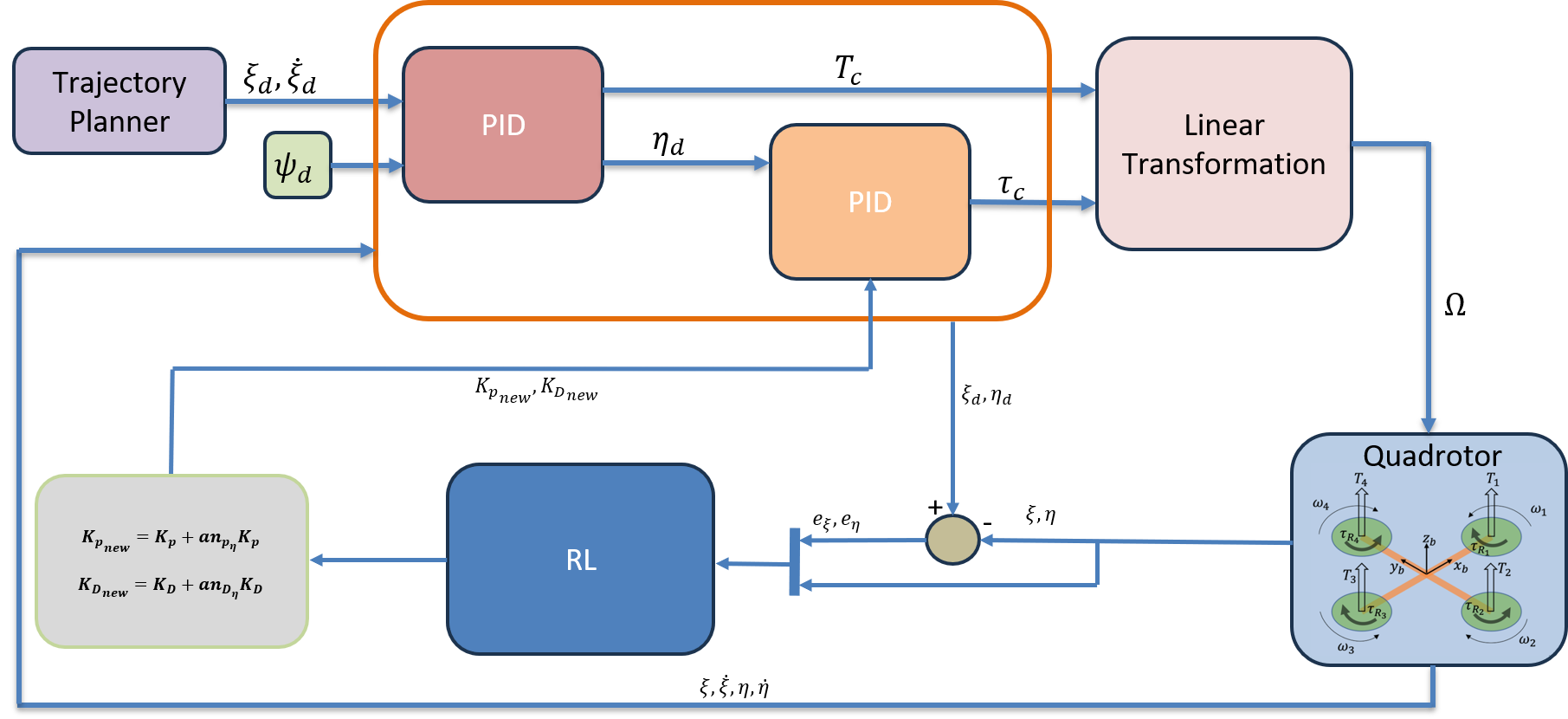}}
\caption{Block diagram of the PD parameter tuning based on RL.}
\label{PID_Tuning}
\end{figure*}

The starting step is the manual tuning of controller parameters. This is followed by employing RL to further refine the values. The block diagram configuration of controller parameters tuning using RL is illustrated in Fig.~\ref{PID_Tuning}. The overall system consists of three components: controller, plant (quadrotor), and agent (RL component). The overall configuration of Fig.~\ref{PID_Tuning} includes a trajectory planner, the linear transformation, and parameter tuning parts. MATLAB/Simulink is utilized to train the agent for the RL-based fine-tuning of the inner loop (attitude) controller. The notation has already been explained.

The state space $S$ (for tuning) includes positions ($p$) along the x, y, and z axes, as well as Euler angles ($\eta$). It also includes position errors ($e_{p}$) along the x, y, and z axes and the errors on the Euler angles ($e_{\eta}$). The agent learns to adjust dynamically the normalized weights ($n_{P_{\psi2}}$, $n_{P_{\varphi\theta1}}$, $n_{P_{\psi1}}$, $n_{P_{\varphi\theta0}}$, $n_{D_{\varphi\theta0}}$), which create the action space ($A \in \mathbb{R}^{5}$) within the range of [-1, 1] for all controller parameters. The state and action spaces are denoted as follows:
\begin{equation}
    \begin{array}{c}
         S = [ p, \eta, e_{p}, e_{\eta} ] \in \mathbb{R}^{12} \\
         A =  [ n_{P1,\varphi \theta}, n_{P1,\psi}, n_{P1,\varphi \theta}, n_{P2,\psi}, n_{D,\varphi \theta} ] \in \mathbb{R}^5
    \end{array}
\end{equation}

Considering the normalized weights, the corresponding parameter tuning equations, related to the gains in \eqref{eq:law_phi} - \eqref{eq:law_phi2} are 

\begin{equation}
\label{eq:gains_modified}
    \begin{cases}
        k_{P1,\varphi \theta,new} = k_{P1,\varphi \theta}(1+ an_{P1,\varphi \theta})\\
        k_{P1,\psi,new} = k_{P1,\psi}(1+an_{P1,\psi})\\
        k_{P2,\varphi \theta,new} = k_{P2,\varphi \theta}(1+an_{P2,\varphi \theta})\\
        k_{P2,\psi,new} = k_{P2,\psi}(1+an_{P2,\psi})\\
        k_{D,\varphi \theta,new} = k_{D,\varphi \theta}(1+an_{D,\varphi \theta})\\
    \end{cases}
\end{equation}
where $a$ represents the search rate and $k_{P1,\varphi \theta,new}$, $k_{P1,\psi,new}$, $k_{P2,\varphi \theta,new}$, $k_{P2,\psi,new}$, and $k_{D,\varphi \theta,new}$ are the new control parameters fine-tuned by the trained RL agent. The manually tuned inner-loop controller parameters serve as the starting gain values for the training while a combination of the search rate and normalized weights contribute to the tuning within an assigned interval relative to the initial controller gains. 

The reward function is defined as a pice-wise function of the attitude error norm,
\begin{equation}
R(||e_\eta||)=
    \begin{cases}
        \ r_1, & \alpha_1 \le ||e_{\eta}||\\
        \ r_2, & \alpha_2 \le ||e_{\eta}|| < \alpha_1 \\
        \ r_3, & \alpha_3 \le ||e_{\eta}|| < \alpha_2 \\
        \ r_4, & \alpha_4 \le ||e_{\eta}|| < \alpha_3 \\
        \ r_5, & \alpha_5 < ||e_{\eta}|| < \alpha_4 \\
        \ r_6, & ||e_{\eta}|| \le \alpha_5 \\
    \end{cases}
\end{equation}
where $r_i$ ($i = 1, \dots,6$) are the reward values, and $\alpha_i$ ($i = 1, \dots,5$) represents the thresholds for the norm of the attitude errors, $||e_{\eta}||$. These thresholds define the conditions under which each reward value, $r_i$, is assigned. 

\begin{figure*}[htb!]
\centerline{\includegraphics[width=1\columnwidth]{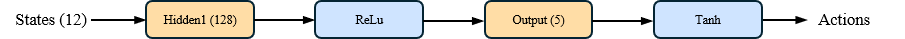}}
\caption{Structure of the actor neural network used for PD parameter tuning.}
\label{Fig:ActorNNTuning}
\end{figure*}

\begin{figure*}[htb!]
\centerline{\includegraphics[width=1\columnwidth]{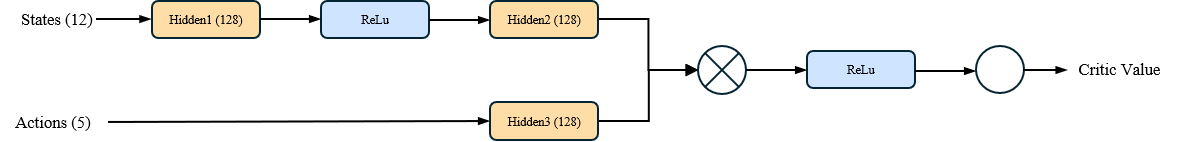}}
\caption{Structure of the critic neural network used for  PD parameter tuning.}
\label{Fig:CriticNNTuning}
\end{figure*}

The actor and critic neural network (NN) are designed as feed forward NNs with structures depicted in Fig.~\ref{Fig:ActorNNTuning} and \ref{Fig:CriticNNTuning}, respectively. The \textit{tanh} activation function is chosen for the actor's output, giving the agent's output the value from the probability distribution between $[-1,1]$.

\section{Agent Environment Setup for Controller Testing} \label{sec4}
The selected target hardware is the Pixhawk 2.1 Cube Black flight control unit which supports PX4-Autopilot firmware.
The controller is developed in MATLAB/Simulink and the deployment process involves overwriting the PX4-Autopilot flight controller with the one resulting from the C++ code generation of the Simulink model. However, challenges arise when certain blocks in the Simulink model are incompatible with the code generation process, as they may not be supported. This limitation can prevent successful code generation, necessitating the modification or replacement of unsupported blocks to achieve a functional executable.

\begin{figure}[hbt]
    \centering
    \includegraphics[width=1\textwidth]{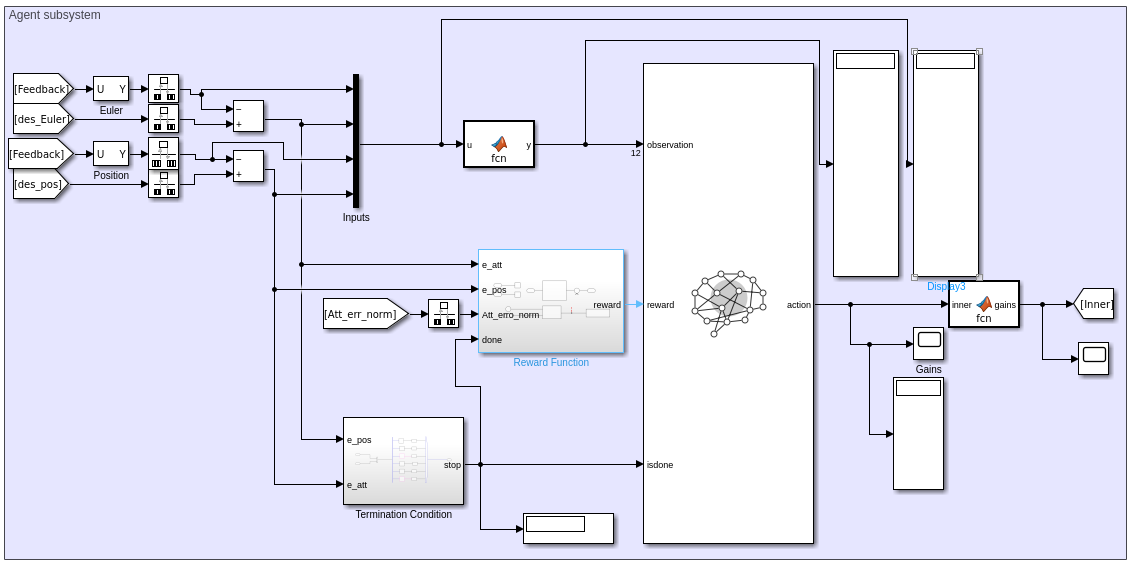}
    \caption{Reinforcement learning Fine-Tuning agent subsystem}
    \label{fig:agent_model_agent}
    \end{figure}

For the proposed controller, the RL agent Simulink block in Fig.~\ref{fig:agent_model_agent}, is not supported for code generation.
To overcome these limitations an alternative approach is implemented: the weights and biases are extracted from the trained action model and the NN is reconstructed within the Simulink workspace. These trained weights and biases, represented as large matrices, are used to replicate the hidden layers of the action model. By applying the same series of mathematical operations of the hidden layers, and utilizing the same observation function employed during the training phase, the original action can be identically reconstructed for the code generation.

In a reinforcement learning agent, the biases and weights within the neural network are fundamental components that allow the model to approximate intricate functions, such as policies or value functions. These parameters are iteratively updated during the training process to optimize the network's performance. The adjustments aim to reduce prediction errors while improving the agent's ability to maximize cumulative rewards, enabling it to learn effective strategies for decision-making in complex environments. To this end,
a neural network must have a sufficient number of neurons to handle complex trajectories effectively. However, in order to deploy the RL agent directly on the Pixhawk flight control unit, the memory limitations of the target hardware need to be taken into account.
To balance performance and memory constraints, a neural network configuration of $128\times128$ neurons is selected. This setup provides satisfactory results while ensuring fesible memory utilization on the hardware.
\subsection{Reconstruction of the Action Layer}
The $tanh$ activation function is applied to the output of the hidden layers in the network, as represented by the equation:

\begin{equation}
     hiddenLayerOutput = tanh(weightsLayer * observation + biasesLayer)
\end{equation}

where $weightsLayer$ and $biasesLayer$ are the parameters specific to this layer, as the equation operates directly on the input observation. These parameters are responsible for determining the transformation of the input data before the nonlinear activation is applied.

For the final layer, the network uses a clipped ReLU function, defined as:
\begin{equation}
    action = min(N,max(Q, weightsLayer * hiddenLayerOutput + biasesLayer))
    \label{eq:action}
\end{equation}
where $N$ and $Q$ are the maximum and minimum allowable action values, respectively. These bounds are determined during the agent's training phase to ensure the output remains within a feasible range. In the case of RL-based fine-tuning of PD controller gains, the values of $N$ and $Q$ are set to $1$ and $-1$, respectively, reflecting the permissible range for the controller's actions. This setup ensures that the network's outputs are appropriately scaled for the control task.

\section{Results} \label{sec5}
\label{sec: combining}

In this research, the focus is set to tracking problem of a circular trajectory, also considering take-off and landing tasks. The trajectory is divided as follows: $10s$ take-off, $2.5s$ hovering, $20s$ circumference (one lap), $2.5s$ hovering, and $10s$ landing for a total of $45s$. The real quadrotor's parameters are manually measured and the motors' characteristic are computed from the relative data-sheet. The resulting values are shown in Table~\ref{tab:param} and used during training and numerical simulations. 

\begin{table}[h]
\centering
\caption{Quadcopter measured parameters}
\label{tab:param}
\begin{tabular}{c|c|c}
\textbf{Parameter} & \textbf{\ \ \ Value\ \ \ } & \textbf{Unit measurement} \\
\hline
$m_{tot}$ & $1.2$ &  kg\\
$m_{prop}$ & $0.01$ & kg\\
$m_m$ & $0.045$ & kg \\
$m_{cg}$ & $0.98$ & kg \\
$r_{cg}$ & $0.0625$ & m \\
$h_{cg}$ & $0.13$ & m \\
$r_m$ & $0.015$ &  m\\
$h_m$ & $0.45$ &  m\\
$r$ & $0.125$ &  m\\
$l$ & $0.225$ &  m\\
$T$ & $8.43$ &  N\\
$\tau$ & $0.1056$ &  $Nm$\\
$I_{xx}$ & $0.0131$ &  $kg m^2$\\
$I_{yy}$ & $0.0131$ &  $kg m^2$\\
$I_{zz}$ & $0.0234$ &  $kg m^2$\\
$k_{motor}$ & $1.4422\times10^{-3}$ &  $\frac{kg m}{rad^2}$\\
$\frac{c_D}{c_L}$ & $0.0237$ &  -\\
$b$ & $3.1427\times10^{-7}$ &  $\frac{kg m}{rad^2}$\\
\end{tabular}
\end{table} 

\subsection{Agent Training Phase}
\begin{figure}[hbt]
\centering
\includegraphics[width=0.7\textwidth]{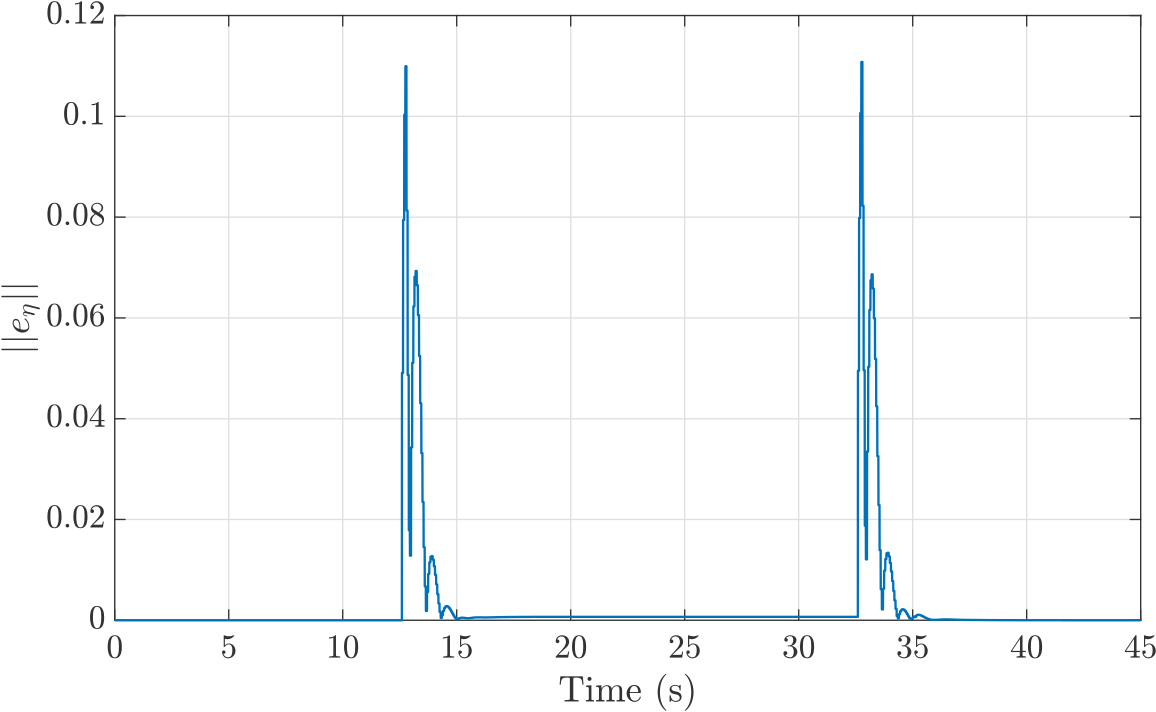}
\caption{Attitude error norm of the manually tuned controller framework.}
\label{fig:att_norm_noagent}
\end{figure}

Before starting the RL training phase, the flight controller is manually tuned to achieve satisfactory tracking performance. After an iterative trial and error tuning process, the controller gain values of Table \ref{tab:gains} are found and the resulting numerical simulation attitude error norm is computed. The plot of the attitude error norm in Fig.~\ref{fig:att_norm_noagent} highlights two main areas with higher spikes in correspondence to the quadrotor's change of direction at $12.5s$ and $32.5s$, hence, at the beginning and the end of the circular trajectory. The training of the agent aims to reduce these high spikes in the attitude error norm.
\begin{table}[h]
    \caption{Gains Table}
    \label{tab:gains}
    \centering
    \begin{tabular}{l|l|l}
\textbf{\textit{Gain}} & \textbf{\textit{Value}}                                                 & \textbf{\textit{Name}}                                      \\ \hline
$k_{P1,z}$            & $8.9$               & Position proportional gain on z-axis     \\
$k_{P2,z}$            & $19.8$              & Velocity proportional gain on z-axis     \\
$k_{P1,xy}$       & $0.6$  & Position proportional gain on $xy$-axis \\
$k_{P2,xy}$       & $3.9$  & Velocity proportional gain on  $xy$-axis \\
$k_{D,xy}$       & $0.29$  & Velocity derivative gain on  $xy$-axis \\
\hline
$k_{P1,\psi}$         & $2$  & Attitude proportional gain for yaw angle     \\
$k_{P2,\psi}$         & $5.4801$ & Angular rate proportional gain for yaw angle \\
$k_{P1,\varphi,\theta}$   & $4$  & Attitude proportional gain for roll and pitch angles     \\
$k_{P2,\varphi,\theta}$   & $11.467$  & Angular rate proportional gain for roll and pitch angles      \\
$k_{D,\varphi,\theta}$   & $0.81905$       & Angular rate derivative gain for roll and pitch angles                        \\
\end{tabular}
\end{table}

According to the proposed methodology of Section \ref{sec3}, RL is then applied to further fine-tune the inner loop controller gains. By heuristically assigning a value of $a=0.4$ to \eqref{eq:gains_modified}, the following updating law is selected
\begin{equation}
\label{eq:gains_modified2}
    \begin{cases}
        k_{P1,\varphi \theta,new} = k_{P1,\varphi \theta}(1+ 0.4n_{P1,\varphi \theta})\\
        k_{P1,\psi,new} = k_{P1,\psi}(1+0.4n_{P1,\psi})\\
        k_{P2,\varphi \theta,new} = k_{P2,\varphi \theta}(1+0.4n_{P2,\varphi \theta})\\
        k_{P2,\psi,new} = k_{P2,\psi}(1+0.4n_{P2,\psi})\\
        k_{D,\varphi \theta,new} = k_{D,\varphi \theta}(1+0.4n_{D,\varphi \theta})\\
    \end{cases}
\end{equation}
The reward function is defined as a function of the attitude error norm. Considering simulation studies after manually tuned controller gains, it is observed that the norm of the attitude error ranges from $10^{-4}$ to $1.1\times10^{-1}$ rad as can be observed in Fig. \ref{fig:att_norm_noagent}. Given this relatively wide range, the reward function is set as the following piece-wise function of the attitude error 
\begin{equation}
R(||e_\eta||)=
    \begin{cases}
        \ -25, & 0.04 \le ||e_{\eta}|| \\
        \ -15, & 0.01 \le ||e_{\eta}|| < 0.04 \\
        \ -10, & 0.001 \le ||e_{\eta}|| < 0.01 \\
        \ \ -5, & 0.0005 \le ||e_{\eta}|| < 0.001 \\
        \ \ -1, & 0.0001 \le ||e_{\eta}|| < 0.0005 \\
        \ \ \ 10, & ||e_{\eta}|| \le 0.0001 \\
    \end{cases} \label{tuning rewards}
\end{equation}

The training phase is carried on in Matlab/Simulink using the Deep Learning and Reinforcement Learning Toolbox, the simulation plant is modelled according to \eqref{FullNEMotionEquation2}-\eqref{FullNEMotionEquation4}, and the selected training hyperparameters are found heuristically and displayed in Table~\ref{tab:TrainingParams}. 
The criterion to complete training is based on achieving a target average reward which value is determined by analyzing the distribution of step counts across the error intervals shown in \eqref{tuning rewards} relative to the norm of the attitude error observed with the manually tuned parameters. 
\begin{table}[h!]
    \caption{Hyperparameters}
    \label{tab:TrainingParams}
    \centering
    \begin{tabular}{l|l}
        \textbf{Parameters} & \textbf{Tuning} \\
        \hline
        Sampling time                   & $0.05$  \\   
        Reward discount factor $\gamma$ & $0.99$  \\   
        Learning rate for actor         & $10^{-3}$ \\
        Learning rate for critic        & $10^{-3}$ \\
        $L_2$ Regularization factor     & $10^{-5}$ \\
        Optimizer parameter $\epsilon$  & $10^{-8}$ \\
        Minimum batch size              & $1024$  \\
        Experience buffer length        & $10^{6}$  \\
    \end{tabular}
\end{table}

According to \eqref{tuning rewards} and the selected sampling time, the step count distribution yields the following values: \([28, 24, 41, 350, 143, 345]\). The baseline reward value obtained from the piecewise function \eqref{tuning rewards} can be calculated as:  
\[
-25 \cdot 28 - 15 \cdot 24 - 10 \cdot 41 - 5 \cdot 350 - 1 \cdot 143 + 10 \cdot 345 = 117.
\]

This reward value serves as a reference point. To achieve better results, an improved reward value is required, necessitating adjustments to surpass this baseline value and drive enhanced performance in the training process.

\subsection{Simulation Results}\label{sec:train_results}
\begin{figure}[hbt]
\centering
\includegraphics[width=0.7\textwidth]{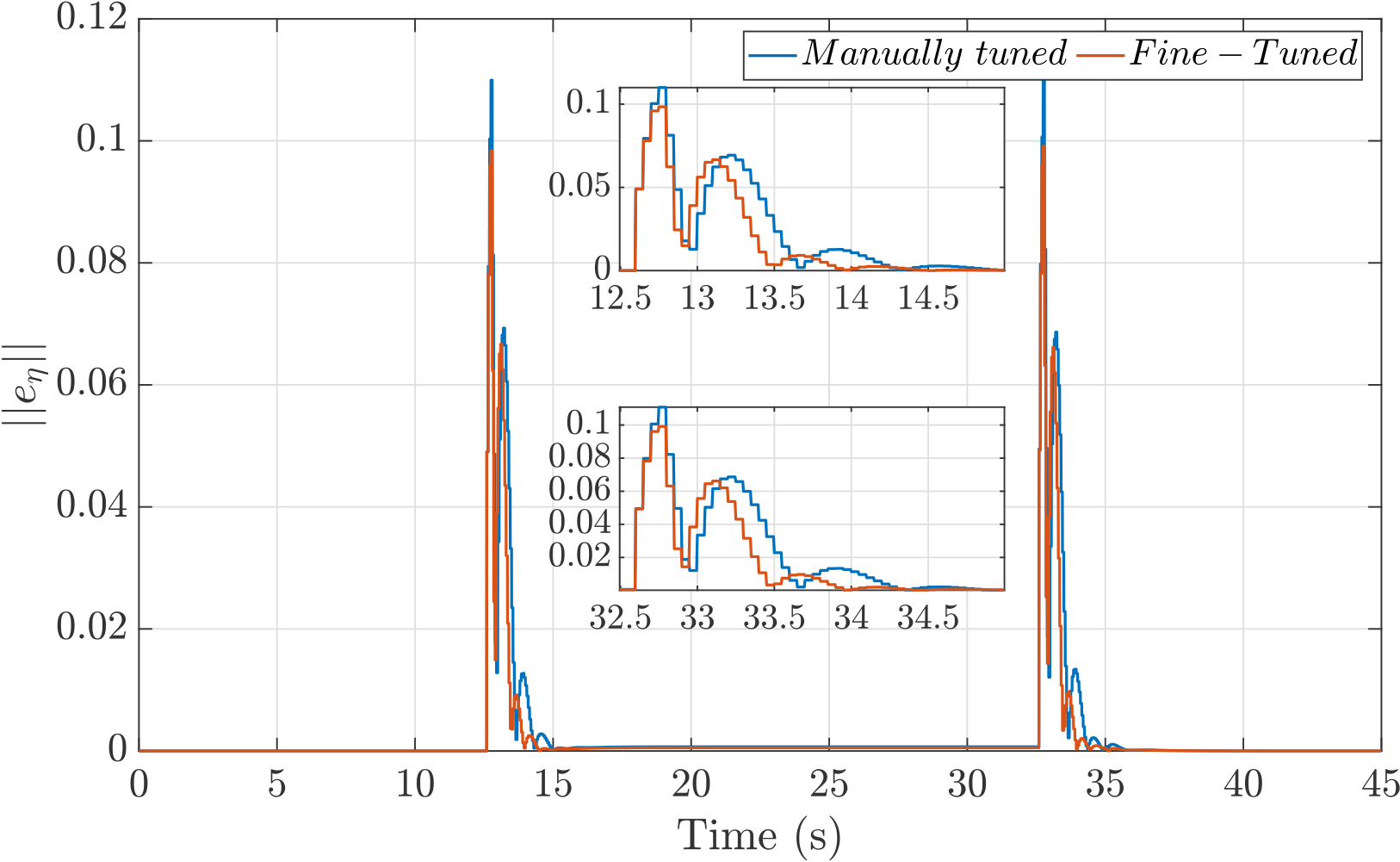}
\caption{Attitude error norms of manually and RL-based fine-tuned controller frameworks.}
\label{fig:att_norm_agent_long}
\end{figure}

Once the training phase is completed, the RL-agent-based controller is tested through numerical simulations. The attitude error norms for both the manually tuned controller and the RL-based fine-tuned controller are presented in Fig.~\ref{fig:att_norm_agent_long}. The results demonstrate that the application of the fine-tuning method reduces overshoots and peak errors while enabling the system to reach steady-state conditions more quickly.
This improvement is attributed to the RL agent, which optimizes the controller by generating a refined set of gains compared to the manually tuned values. These adjustments result in enhanced performance and stability of the control framework. Table~\ref{table:RMSE_Sim} presents the RMSE values for attitude errors norm, providing a clear and quantitative comparison between the performance of the manually tuned controller and the RL fine-tuned one. The fine-tuned gains result in better signal tracking and an overall improved system response compared to the manually tuned gains, demonstrating the effectiveness of the RL-based approach.

\begin{table}[h!]
\centering
\caption{Numerical Simulation RMSE of Attitude Error Norm}
\label{table:RMSE_Sim}
\begin{tabular}{c|c|c}
\textbf{Numerical Simulation} & \textbf{ Manually tuned} & \textbf{\ \ \ Fine-Tuned\ \ \ } \\
\hline
$|e_{\eta}|_{RMSE}$ & $12.75\cdot 10^{-3} $ &  $11.17\cdot 10^{-3}$ 
\end{tabular}
\end{table}

\begin{figure}[hbt]
\centering
\includegraphics[width=0.7\textwidth]{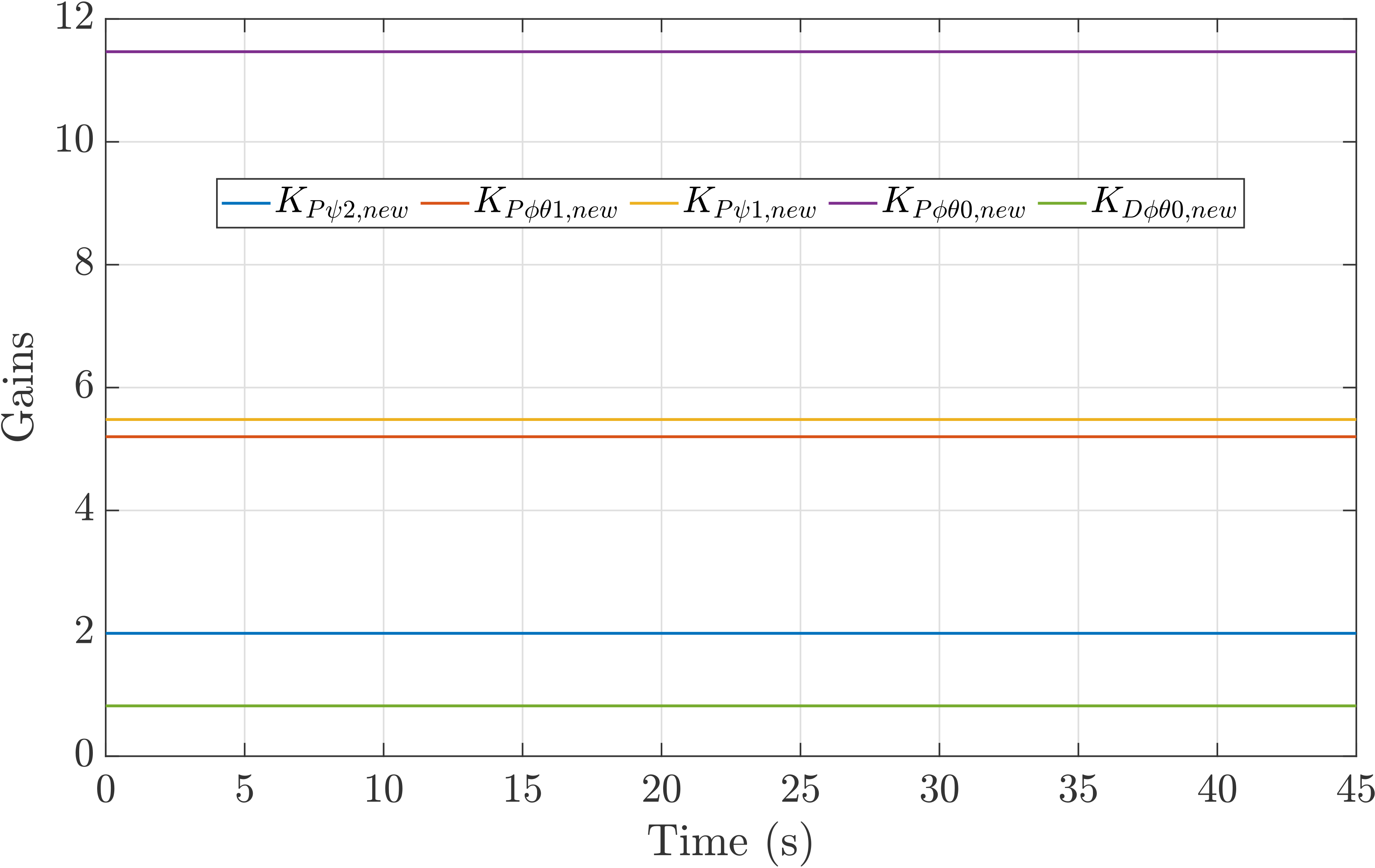}
\caption{The inner loop controller gains of RL-based fine-tuned controller framework in time.}
\label{fig:gains3_sitl}
\end{figure}

Additionally, as shown in Fig.~\ref{fig:gains3_sitl}, the RL agent sets the fine-tuned controller gains differently from the initial manually tuned values right from the start of the simulation. Given the lack of disturbances in the numerical simulations, the fine-tuned gains remain constant and bounded throughout the simulation.

\subsection{Experimental results}
The proposed methodology is deployed on quadrotor hardware to further test whether the RL-agent-based controller performance improvements are carried out in real outdoor flight conditions. To this end, the C++ code was uploaded to the Pixhawk 2.1 Cube Black flight control unit following the procedure in Section (\ref{sec4}).

\begin{figure}[hbt]
\centering
\includegraphics[width=0.7\textwidth]{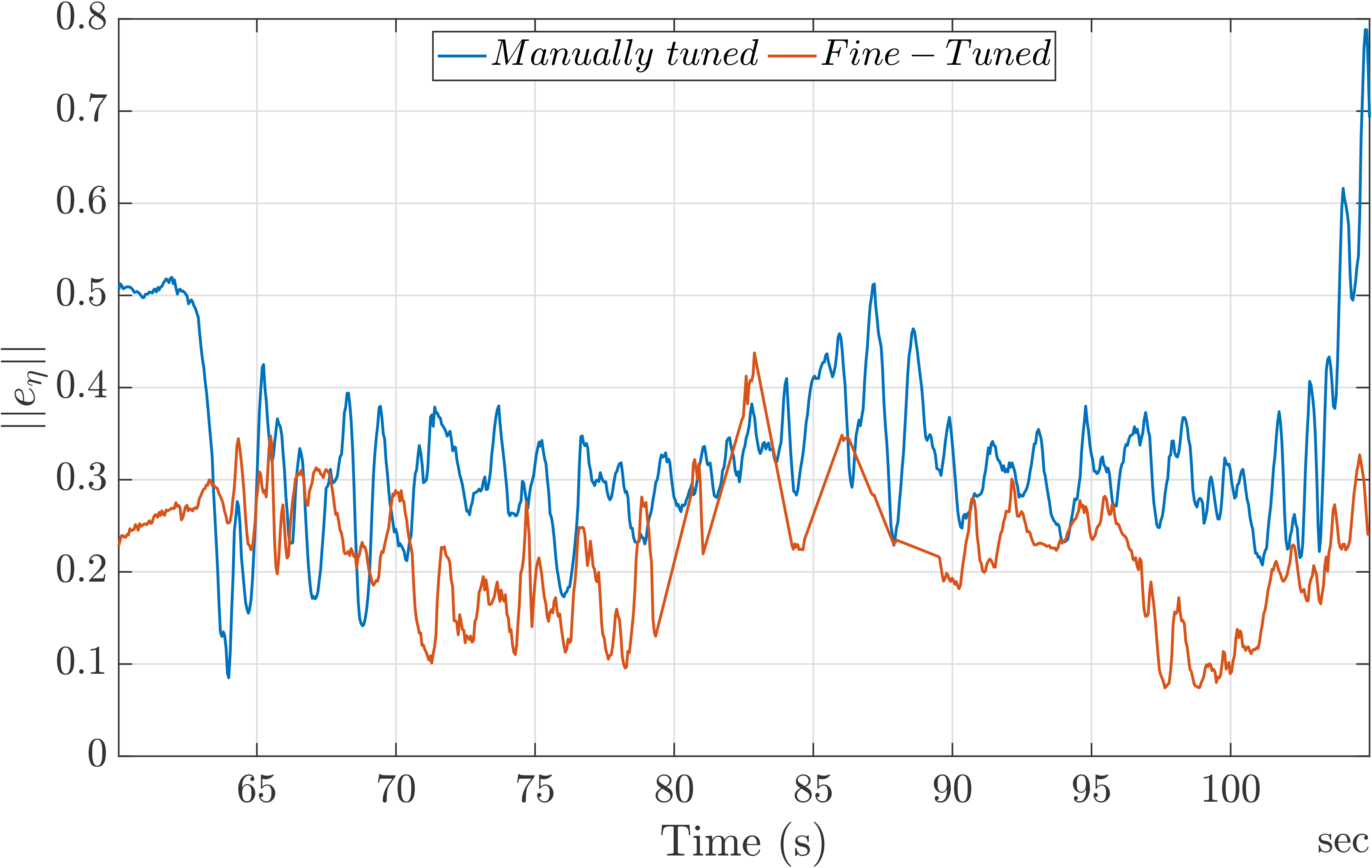}
\caption{Attitude error norms of manually and RL-based fine-tuned controller frameworks for the experimental simulation.}
\label{fig:att_err_norm_out}
\end{figure}

As in the previous section, performances are compared in terms of the norm of attitude errors. Fig.~\ref{fig:att_err_norm_out} compares the performance of the manually tuned gains with that of the fine-tuned gains, highlighting the differences in error magnitudes and response characteristics between the two approaches.


Overall, the observed behavior for the trajectory shows noticeable performance enhancement, with the fine-tuned gains outperforming the manually tuned ones. This highlights the RL agent's capability to adapt and optimize control performance, even in the presence of real world disturbances. This results can be further verified by comparing the respective RMSE value presented in Tab. \ref{table:RMSE_Out}.
\begin{table}[h!]
\centering
\caption{Numerical Simulation RMSE of Attitude Error Norm}
\label{table:RMSE_Out}
\begin{tabular}{c|c|c}
\textbf{Outdoor Flight} & \textbf{ Manually tuned} & \textbf{\ \ \ Fine-Tuned\ \ \ } \\
\hline
$|e_{\eta}|_{RMSE}$ & $33.93\cdot 10^{-2}$ &  $22.55\cdot 10^{-2}$
\end{tabular}
\end{table}

\begin{figure}[hbt]
    \centering
    \includegraphics[width=0.7\textwidth]{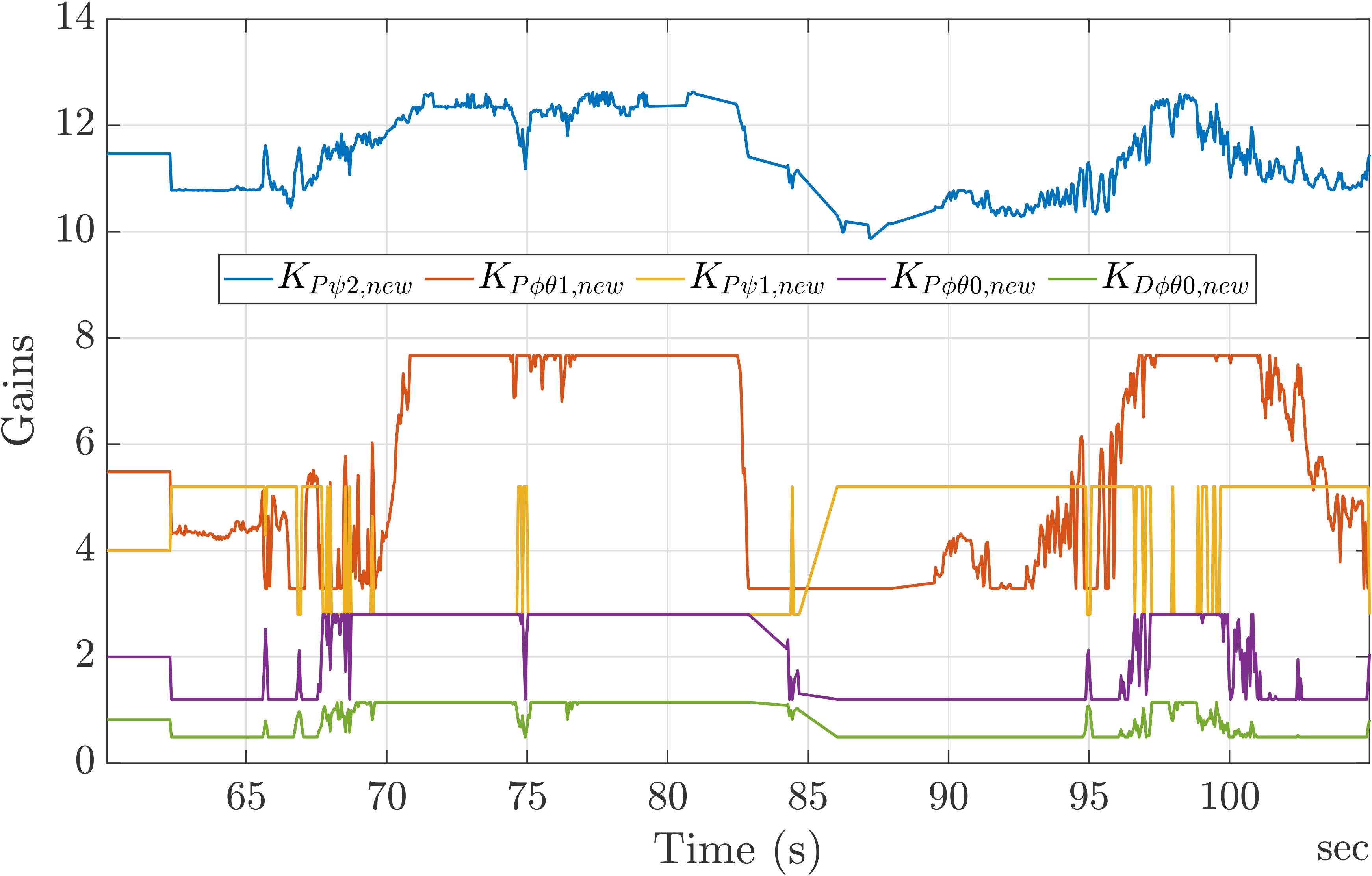}
    \caption{Time varying gains in outdoor experimental test results}
    \label{fig:gains_circ2_out}
    \end{figure}

As a further impressive result, the RL-agent-based controller is able to adapt online the controller gains. This behavior shows that, even if not necessary during numerical simulation ideal conditions, the agent has learned how to adapt the controller parameters in case of disturbances. The evolution of the RL-fine-tuned controller gains is shown in Fig.~\ref{fig:gains_circ2_out}, and, although initialized to match the manually tuned values, after approximately $2s$, the gains begin to adjust dynamically. These changes are moderate, avoiding extreme values, and demonstrate the agent's ability to refine the control parameters in response to the system's requirements without overextending them. This behavior reflects a balanced adaptation aimed at maintaining system stability and performance.
The agent operates without prior knowledge of the outdoor environment and must contend with external factors such as wind gusts and ground effects. Throughout the flight test, the agent's adaptive behavior is evident as it attempts to mitigate these disturbances. In some instances, the gains reach their maximum or minimum allowable values, reflecting the system's efforts to stabilize under challenging conditions. Additionally, a few sudden spikes in the gains are observed, indicating rapid adjustments made by the agent to counteract abrupt changes in the environment. This behavior highlights the agent's responsiveness and the challenges posed by dynamic and unpredictable external forces.
Overall the application of the RL strategy improved the performances of flight outmatching those of the manually tuned gains.

\section{Discussion and Conclusion}\label{sec6}

The presented results highlight both the potential and the challenges of applying reinforcement learning (RL)-based fine-tuning for PD controllers in real-world scenarios. One of the main issues observed, particularly in the outdoor tests, was the significant offset introduced by GPS inaccuracies. This initial error can degrade the overall performance of the flight, as the controller must compensate for this bias while maintaining trajectory tracking. Despite this limitation, the RL-based controller demonstrated superior performance compared to the manually tuned controller, showcasing its ability to adapt and optimize under challenging conditions.

The DDPG-based RL algorithm, an off-policy actor-critic approach, proved effective for fine-tuning the inner-loop gains of the controller. By leveraging this methodology, a robust solution was achieved, capable of handling complex trajectories involving both circular paths and hovering segments, which mirror realistic flight scenarios. It is worth noting that drag and gyroscopic effects were omitted during the training phase but reintroduced in subsequent tests. This approach allowed for a controlled evaluation of the RL agent’s adaptability to real-world conditions. Despite these simplifications during training, the RL-enhanced controller consistently outperformed the manually tuned one in simulated, Hardware-In-The-Loop (HIL), and outdoor tests.

This study successfully demonstrated the viability and effectiveness of a DDPG-based RL algorithm for fine-tuning PD controller parameters. The proposed method achieved substantial performance improvements across various test environments, including realistic outdoor scenarios. The RL-tuned controller exhibited enhanced adaptability, precise trajectory tracking, and robustness in the face of environmental disturbances. These results underscore the potential of RL-based approaches to bridge the gap between simulation and real-world applications in UAV control.

While the results are promising, several areas for future research remain. First, addressing GPS-related issues is critical for improving overall system accuracy. Integrating more precise positioning systems, such as RTK GPS, or exploring GPS-independent navigation methods using optical flow cameras could significantly reduce initial offsets and enhance performance.

Additionally, incorporating disturbances such as wind gusts, ground effects, and sensor noise directly into the training phase could further improve the agent's adaptability to real-world conditions. Such an approach would create a more realistic simulation environment, allowing the RL agent to develop strategies that account for these factors from the outset. This enhanced training process could lead to even greater error reduction and better performance in outdoor scenarios.

The findings of this study highlight the transformative potential of RL-based fine-tuning for UAV control, particularly in scenarios where traditional methods face limitations. The demonstrated adaptability of the RL-tuned controller to real-world disturbances, coupled with its ability to optimize performance across diverse conditions, positions this approach as a practical and scalable solution for modern UAV applications. The proposed methodology bridges the gap between theoretical development and real-world deployment by systematically addressing challenges such as initial positioning errors and unmodeled disturbances. This approach is recommended not only for its robustness and precision but also for its capacity to handle dynamic and unpredictable environments, making it a strong candidate for future UAV systems that demand reliability and flexibility.

Finally, extending this study to include more complex trajectories, diverse flight conditions, and multi-agent coordination could provide valuable insights into the scalability and versatility of RL-based control strategies. These advancements would further solidify RL as a powerful tool for UAV control in dynamic and unpredictable environments.

\section*{Acknowledgments}
This research has been partially supported by the Ministry of National Education of the Republic of Turkey on behalf of the Istanbul Medeniyet University, Turkey, and the D. F. Ritchie School of Engineering and Computer Science, University of Denver, CO 80208.

\bibliographystyle{unsrt}  
\bibliography{templateArxiv}

\end{document}